\documentstyle[aps,prl,epsf]{revtex}
\begin{document}
\draft

\title{The 2-d Coulomb Gas on a 1-d lattice}
\author{Onuttom Narayan$^1$ and B. Sriram Shastry$^2$}
\address{$^1$ Physics Department, University of California, Santa Cruz, 
CA 95064, USA}
\address{$^2$ Physics Department, Indian Institute of Science, 
Bangalore 560012, INDIA}
\date{\today}
\maketitle

\begin{abstract}
\qquad The statistical mechanics of a two dimensional Coulomb gas confined to
one dimension is studied, wherein hard core particles move on a ring.
Exact self-duality is shown for a version of the sine-Gordon model
arising in this context, thereby locating the transition temperature exactly.
We present asymptotically exact results for the correlations in the
model and
characterize the low and high temperature phases. Numerical simulations
provide support to these renormalization group calculations. Connections
with other interesting problems, the quantum Brownian motion of a particle
in a periodic potential and impurity problems, are pointed out.
\end{abstract}

$\overline{}$

\section{Introduction}

The physics of the inverse squared exchange Heisenberg model, the
so called Haldane Shastry model \cite{hal,shas} has been a field
of considerable activity recently. Here one considers a spin ${1\over 2}$
Heisenberg antiferromagnet in 1-dimension (1-d) on a ring of L sites
with an exchange Hamiltonian 
\begin{equation} H=J\sum_{i<j}\frac{\phi^{2}}
{\sin (\phi (x_{i}-x_{j}))^{2}}\;\vec{S_{i}} \cdot\vec{S_{j}}
\label{ham} 
\end{equation} 
with $x_{j}$ as integers denoting lattice points and 
$\phi =\pi/L$. The ground state of the model of Eq.(\ref{ham}) in a sector 
with $N$ spin reversals (relative to the ferromagnet) located at
$\{..x_{j}..\}$ is given by 
\begin{equation} 
\psi (x_{1},x_{2},\ldots,x_{N})=(-1)^{\sum x_{j}}
\prod_{i<j}\sin ^{\beta /2}(\phi (x_{j}-x_{i}))
\label{wvfn} \\ 
\end{equation} 
with $\beta =4.$ The model is interesting from several points of view,
such as the connection with Gutzwiller projection in strongly correlated
systems, and from the intimate connection with the isotropic Heisenberg
antiferromagnet, the Bethe chain. The spin- spin correlation function of
the above wave-function at $\beta =4$ has the same decay exponent as the Bethe
chain, namely unity. Further, at $\beta =2,$ the wave function is the
ground state of the free Fermi gas, with either long ranged hops or just
nearest neighbor hops. The first case at $\beta=2$ 
corresponds to dropping the $zz$
part of Eq.(\ref{ham}), and the second to the anisotropic Heisenberg
model: $H=\sum_{i}(S_{i}^{x}S_{i+1}^{x}+S_{i}^{y}S_{i+1}^{y}+\Delta
S_{i}^{z}S_{i+1}^{z})$ at $\Delta =0.$ Thus we find that $\beta =4$ and
2 are in close correspondence with $\Delta =1$ and 0 respectively. The
Heisenberg model is well known\cite{baxter} to have a transition to a
massive N\`{e}el ordered phase at $\Delta =1$, and so one might suspect
that the wave function Eq.(\ref{wvfn}) could develop long ranged order as
$\beta $ increases from 4, perhaps even at $\beta =4^{+},$ a possibility
we shall investigate in this paper. It is obvious that N\`{e}el order can
also be viewed as crystalline order of hard core bosons, where the bosons
correspond to the spin reversals of the Heisenberg system via the familiar
lattice gas analogy. We will almost exclusively use this point of view
below. Note that the density variables $\rho _{i}=(\frac{1}{2}-S_{i}^{z})$
take values 0,1 so that we can map density correlators to spins readily,
with $N$ the number of hard core particles restricted to $N\leq L/2$.

For the case of $\beta \neq 2,4$ the wave functions do not represent
either symmetric or antisymmetric functions, and are hard to interpret
as physically allowed states for bosons/fermions, unless one imposes a
rather non-analytic restriction of taking the modulus. For larger even
integer values of $\beta =6,8..$ the wave function Eq.(\ref{wvfn}) is an
eigenstate of the anisotropic version of the Hamiltonian Eq.(\ref{ham}),
but only in restricted sectors of numbers of particles, for fillings
up to $\frac{2}{\beta}$, since beyond this filling the states are no
longer ``good functions'' in the sense of Ref.~\cite{shas2}, {\it i.e.\/}
they have Fourier components that ``spill out'' of the first Brillouin 
zone, requiring umklapp.

The evaluation of correlations in the above wavefunction reduces to those
of a 2-d Coulomb gas confined to a 1-d ring, but with the positions of the
particles discretized to a lattice. This is a far reaching distinction
from the case where the charged particles are in the continuum, a
case that is familiar from the well known results of Dyson, Mehta and
Gaudin\cite{mehta1} for random matrices. In the latter case, the Coulomb
gas does not crystallize in the sense of possessing {\it LRO, }although
the density correlators are arbitrarily slowly decaying. In the lattice
case one expects {\it LRO, } which is consistent with N\`{e}el order.

The discrete Coulomb gas has been subject to a few exact calculations
earlier. Gaudin \cite{gaudin} computed the normalization constant of the
wavefunction and the grand partition function exactly at three values of
$\beta =1,2,4.$ His isothermal calculation of the grand partition function
at these values of the temperature gives the distribution of zeroes in
the thermodynamic limit as lying on segments of the unit circle. Mehta
and Mehta \cite{mehta2} computed the density correlators exactly at these
values of $\beta.$ The calculations at $\beta =2$ are not unexpected,
since the model at $\beta=2$
reduces to a free Fermi lattice-gas, but the other cases are
highly non trivial. Sutherland \cite{sutherland} has presented results
at zero temperature for the allowed ground state patterns, that turn
out to be quite complex for arbitrary rational fillings.

In this work, we present some asymptotically exact results on
this problem, using a combination of renormalization group, and exact
duality arguments on related models. We only consider simple rational
fillings in this work with filling $f=\frac{N}{L}=1/2,1/3,1/4...$ We find
that for each such filling f there is a transition temperature $\beta
_{c}=2/f^{2}$ at which the {\it LRO\/} sets in. In Section 2 of this paper,
we show that the discrete Coulomb gas (DCG) is asymptotically equivalent
to a sine Gordon model that has been studied extensively in connection
with several interesting problems recently. This is achieved through a
series of approximations that lead to a phonon representation at
high temperatures, and a kink representation at low temperatures.
Correlation functions are discussed in both representations. In Section 3,
the analogs of the phonon and kink representations are constructed for
a model asymptotically equivalent to the discrete Coulomb gas, and an
exact duality connecting the two pictures is obtained.
The duality found by us is closely related to that found by Kjaer and
Hilhorst\cite{kjaer}, who studied  a discrete height problem- a roughening
model, where the heights interact via  a $1/r^2$ interaction. Our sine Gordon
model reduces to this model on integrating out the Gaussian displacements.
Section 4 presents
numerical results that confirm the results of the previous sections and
the  difficulties involved in extracting true exponents are highlighted.
Connections to related models and other general issues are discussed in
Section 5.

\section{ Phonon and kink representations}

In order to understand the spin correlations for these wavefunctions,
it is convenient to convert the problem into one in classical
statistical mechanics. For a ring with $N=f\;L\;$ spins, if 
$\psi(x_{1},x_{2}\ldots x_{N})$ is the amplitude for the down spins
located at $x_{1},x_{2}\ldots x_{N},$ then 
$|\psi (x_{1},x_{2},\ldots x_{N})|^{2}$ is the corresponding
probability. By rotational symmetry, the spin-spin correlation
function $ \langle {\bf S}(x)\cdot {\bf S}(y)\rangle $ is equal to
$3S_{z}(x)S_{z}(y).$ This can be calculated from $|\psi |^{2},$ without
any knowledge of the phase of $\psi .$ (This is not possible in general
for higher order correlations, where one cannot always get rid of $S_{+}$
and $S_{-}$ operators by symmetry arguments. In this paper, we shall
only consider two-point correlations.)

If we express $|\psi (x_{1},x_{2}\ldots x_{N})|^{2}$ as 
$\exp [\ln |\psi|^{2}],$ we can view $-\ln |\psi |^{2}$ as the energy of
a classical system of $N$ particles (distributed over $L=N/f$ sites), and
$|\psi |^{2}$ as the statistical weight assigned in thermal equilibrium to
a particle configuration. For the wavefunctions we consider here, 
$-\ln|\psi |^{2}$ has the form 
\begin{equation} 
-\ln |\psi(x_{1},x_{2}\ldots x_{N})|^{2}=-{\frac{\beta }{2}}
\sum_{i<j}\ln \big[ \phi ^{-2}\sin ^{2}\phi {(x_{i}-x_{j})}\big]
\qquad +{\rm const}.
\label{energy} 
\end{equation} 
The additive constant at the end of the right hand side is necessary
in order to ensure that the wavefunction is normalized; however, in
the statistical mechanics picture, it only gives rise to an overall
proportionality factor in the partition function, and can be ignored. The
$1/\phi^2$ inside the argument of the logarithm has been pulled out from
the additive constant so that the $L\rightarrow \infty $ limit exists.

Eq.(\ref{energy}) describes a collection of particles with pairwise
interactions. Every particle repels every other particle logarithmically.
The argument of the logarithm is effectively (the square of) the
straight line (or chord) distance between two points $x_{i}$ and
$x_{j}$ on the ring.  Thus the system consists of a charged gas with
two-dimensional Coulomb interactions, but with the particles confined to
a (one-dimensional) ring lattice. The prefactor $\beta $ has a natural
interpretation of the inverse temperature.

In its ground state, the system has one particle on every $1/f^{th}$
site, at least for simple fractions $f$ that we consider here ( say
1/2, 1/3 etc.). As the temperature is lowered, {\it i.e.\/} $\beta $
is raised, there is the prospect of the system crystallizing into a
long-range ordered state.  As we shall see in this paper, this indeed
happens at $\beta =2/f^{2},$ and is the result of a combination of two
factors. Firstly, although one normally expects short range order in a
one-dimensional system, the long ranged logarithmic interactions convert
this to quasi long range order even at high temperatures. Secondly,
the restriction that particles can only be placed on lattice sites
crystallizes the system at low temperatures.

There are two complementary approximations that one can make on the
ring-lattice Coulomb gas, one appropriate for high temperatures and one
for low temperatures. Both these approximations yield a one-dimensional
long-ranged sine-Gordon model, but with different parameters, reflecting
the well known duality of this model. This duality is usually derived
for a continuum sine-Gordon model,\cite{schmid}
but there are slight differences for
the lattice version, as we shall now see.

\subsection{\noindent Phonon representation}

At high temperatures, there are large fluctuations in the separation
between neighboring particles. It is reasonable to expect that,
in this regime, the underlying lattice constraint might not be very
important. Accordingly, in Eq.(\ref{energy}), we express $x_{i}$ $=$
$\frac{1}{f}(i+u_{i}),$ where $u_{i}/f$ is the deviation from an ideal
crystalline state. The hard lattice constraint is replaced with a periodic
potential, to obtain 
\begin{equation} 
\beta H\big[\{u\}\big]=-{\frac{\beta}{2}}\sum_{i<j}^{{}}\ln 
\bigg[ {\frac{{L^{2}}}{{\pi ^{2}}}}\sin^{2}
{\frac{{\pi (i+u_{i}-j-u_{j})}}{f\;{L}}}\bigg]+\sum_{i}V(u_{i}).
\label{htenergy} 
\end{equation} 
The $u_{i}$'s are now continuous variables, and the potential $V$
favors locating the particles at lattice sites. As an example consider
$V(u_{i})=-|const|\cos (2\pi x_{i})$ which reduces to 
$-|const|\cos (2\pi u_{i}/f)$, leading to the sine Gordon theory
considered below. More formally, the hard lattice constraint
can be expressed in terms of a singular periodic potential $V$:
$V(u)=\ln\big[\sum_{n}\exp (2\pi niu/f)\big].$ We shall instead consider the
general class of potentials with periodicity ${f},$ and later exploit the
universality under the renormalization group.  Expanding the first term
on the right hand side of Eq.(\ref{htenergy}), to second order in $u$ we
have 
\begin{equation} 
\beta H\big[\{u\}\big]={\frac{\beta }{4}}{\frac{{\pi^{2}}}{N{^{2}}}}
\sum_{i<j}^{^{{}}}{\frac{{(u_{i}-u_{j})^{2}}}{{\sin^{2}
\big[\pi (i-j)/N\big]}}}+\sum_{i}V(u_{i}).  
\label{phonon1}
\end{equation} 
In Fourier space, expanding $u_{j}$ $=$
$\sum_{q}(\widetilde{u}_{q}/\sqrt{N})\exp [i\;j\;q],$ it can be
shown\footnote{We use ($\frac{\pi }{N})^{2}$
$\sum_{n=1}^{N-1}$cosec$^{2}(\frac{n\pi }{N})\exp (ikn)=
-|k|\pi +\frac{1}{2}k^{2}+\frac{\pi ^{2}}{3}(1-
\frac{1}{N^{2}})$ with $-\pi \leq k\leq \pi $.}
that this is equivalent to 
\begin{equation} 
\beta H\big[\{u\}\big]={\frac{\beta }{2}}\sum_{q}\big(\pi |q|-q^{2}/2\big)
\widetilde{u}_{q}\widetilde{u}_{-q}+\sum_{i}V(u_{i}).  
\label{phonon2} 
\end{equation}
The sum over $q$ ranges from $-\pi $ to $\pi .$ Compared to the $\pi |q|$
term, the $q^{2}$ is irrelevant in the renormalization group sense,
and can be ignored in calculating long wavelength properties. Higher
order terms in the expansion of Eq.(\ref{htenergy}) in powers of $u$
are similarly unimportant, as we see later.

\subsection{Coulomb Gas}

We now cast the problem of computing the partition function for 
Eq.(\ref {phonon2}) in the form of another Coulomb gas,
but with a variable number of charge pairs, controlled
by a chemical potential. Writing a general expansion 
$\exp\big(-V(u_{j})\big)=\sum_{m_{j}}c_{m_{j}}\exp (2\pi iu_{j}m_{j}/f),$
consistent with the periodicity $u\rightarrow u+f$ we consider
the partition function 
\begin{equation} 
Z=\int \![du]\,\exp\Big[-\sum_{q}G(q)\widetilde{u}_{q}\widetilde{u}
 _{-q}\Big]\prod_{j}\Big(\sum_{m_{j}}c_{m_{j}}e^{2\pi iu_{j}m_{j}/f}\Big) 
\label{coleman1}
\end{equation} 
where the $u_{j}$'s are continuous variables, and $G(q)$ is some function
of $q$. (Corresponding to Eq.(\ref{phonon2}), one would have 
$G(q)=(\beta/2)(\pi|q|-q^2/2)$.) This is
equivalent to 
\begin{equation} 
Z=\sum_{m_{1}}\sum_{m_{2}}\ldots \int \![du]\,\exp\Big[-\sum_{q}G(q)
\widetilde{u}_{q}\widetilde{u}_{-q}\Big]\Big(\prod_{j}c_{m_{j}}\Big) 
e^{2\pi i/f\sum_{j}u_{j}m_{j}}.  
\label{coleman2} 
\end{equation} 
If $G(q=0)=0,$ the only terms in this sum that are not zero are
those for which $\sum_{j}m_{j}=0.$ Integrating out the $u$'s,
we have 
\begin{equation} 
Z\propto \sum_{m_{1}}\sum_{m_{2}}\ldots\delta _{\sum m_{j},0}
\Big(\prod_{j}c_{m_{j}}\Big)\exp\Big[-\pi^{2}/f^{2}\sum_{q}
(\widetilde{m}_{q}\widetilde{m}_{-q})/G(q)\Big].
\label{gaspart} 
\end{equation} 
This is the partition function for a charged gas. (If we restrict $c_{m}$
to be non-zero only for $m=\pm 1$ or 0, we have a dilute gas of unit
positive and negative charges.) Using the fact that $\sum_{j}m_{j}=0,$ 
the exponent in the exponential is equal to 
\begin{equation} 
{\frac{{2\pi^{2}}}{Lf^{2}}}^{{}}\sum_{i<j}m_{i}m_{j}
\sum_{q}{\frac{{1-\cos q(i-j)}}{{G(q)}}}=\frac{\pi }{f^{2}}
\sum_{i<j}m_{i}m_{j}\int\!dq\,{\frac{{1-\cos q(i-j)}}{{G(q)}}}  
\label{exponent2} 
\end{equation}
where we have taken the $L\rightarrow \infty $ limit in the last
step. For the case $G(q)=\frac{\pi \beta }{2}|q|$ , i.e. the leading low
energy part of Eq.(\ref{phonon2}), we can evaluate the integral easily
for large separation of the charges, and find the dilute Coulomb gas
partition function:
\begin{equation}
Z_{k-\bar{k}}=\sum_{m_{i}=-\infty }^{\infty }\delta _{\Sigma
m_{i}=0}\Big(\prod_{j}c_{m_{j}}\Big)\exp 
\bigg[ -\beta _{eff}\Big\{-\sum_{i<j}\log |i-j|\;m_{i}m_{j}-\mu _{eff}
\sum m_{i}^{2}\Big\}\bigg]  \label{kinkpart}
\end{equation}
with $\beta _{eff}=\frac{4}{\beta f^{2}}$ and the `chemical potential' 
$\mu_{eff}=-(\gamma +\log \pi )/2=-.860973$ 
with \smallskip $\gamma $ Euler's constant ($.577216$). The Coulomb 
interaction binds unlike charges, and repels like charges. The quadratic
terms of $G(q)$ give rise to a sublogarithmic part in the interaction
energy, but do not affect $\mu _{eff}. $ The object $\mu _{eff}$ is
not really a chemical potential, since the relative weights assigned
to different charges depends on the coefficients $\{c_{m}\}.$ However,
for large $\beta _{eff},$ where the partition function is dominated by
$m_{i}=0,\pm 1,$ $\mu _{eff}$ may be viewed as the chemical potential
associated with the (unit) positive and negative charges in the system,
provided that $c_{\pm 1}=c_{0}.$ Alternatively, $\mu_{eff}$ can be 
absorbed in a redefinition of the coefficients $c_{m},$ and is included 
when we start with general periodic potentials in Eq.(\ref{phonon2}).

\subsection{\noindent Kink representation}

In order to develop this representation, it is convenient to first 
rewrite Eq.(\ref{energy}) as 
\begin{equation}
\beta H\big[\{\rho \}\big]=-\beta f^2\sum_{\{i<j\}}m_{i}m_{j}
\ln \bigg[{L\over\pi}\Big\vert\sin 
{{\pi (i-j)}\over L}\Big\vert\bigg] +{\rm const}  
\label{ltenergy}
\end{equation}
The variable $m_{i}$ is $(1-f)/f$ if the site $i$ is occupied, and
$-1$ if the site is unoccupied. Compared to Eq.(\ref{energy}), this is
equivalent to adding a background charge of $-f$ at every lattice site,
and then reducing the unit of charge to $f$; although this changes the
total energy (the additive constant in Eq.(\ref{ltenergy}) is different
from that in Eq.(\ref {energy})), the energy difference between different
configurations---and therefore their relative statistical weight---is
unaltered.

At low temperatures, the system is in an almost perfect crystalline state,
with one particle after every $1/f$ sites. There are long segments that
are shifted by $l$ sites with respect to a reference crystalline state,
with $l=0,1,2\ldots (1/f-1).$ (Shifting a segment by $1/f$ sites is
equivalent to not shifting it at all.) There is an effective buildup of
charge at the ``kinks'' at the segment boundaries. At low temperatures,
the dominant configurations have neighboring segments with a relative
shift $\pm 1$ corresponding to a unit positive charge (a kink) or a unit
negative charge (an antikink) at the segment boundary. Note that there
is {\it no\/} alternation rule for the kinks here, unlike the case for
a system where the ground state has ferromagnetic order rather than
antiferromagnetic.

\begin{figure}
\centerline{
\epsfxsize\columnwidth\epsfbox{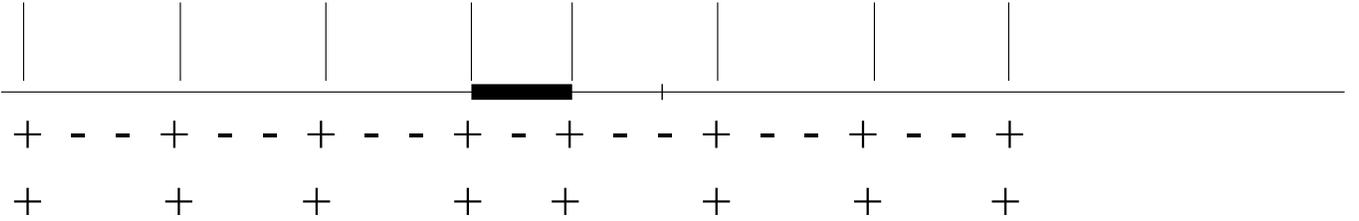}}
\vskip 0.25truecm
\caption{Typical lattice segment for $f=1/3.$ Every positively charged
site is doubly charged. The vertical lines denote the boundaries between
adjacent quadrupoles, and lie on the midpoints of the positively charged
sites. The thick section in the middle of the line segment is a kink,
consisting of one positive charge at either end and only {\it one\/}
negative charge inside. The total charge of the kink is thus $+1.$}
\label{kink}
\end{figure}

We now show that the energy of the system can be understood as an
interaction of these charged kinks. As illustrated in Figure~\ref{kink}, 
two neighboring segments can be viewed as each consisting of a long 
string of quadrupoles, with a residual charge between them. Each 
quadrupole
consists of $f-1$ negative charges terminated by a charge $+(f-1)/2$
at either end.  The residual charge associated with the kink is $\pm 1;$
in Figure~\ref{kink} , it is $+1.$ Since the 
interaction energy between a quadrupole
at $i$ and a charge at $j$ decays as $1/(i-j)^{2},$ the total interaction
energy between {\it all\/} the quadrupoles in a segment and a kink not
at its terminus decays as $1/l$ when the typical segment size is $l.$
The interaction between the quadrupoles in non-adjacent segments can
likewise be neglected. One is left with the interaction between the
kinks. (The interaction between a kink and its adjacent segments, or
two neighboring segments separated by a kink, can be interpreted as
a self-energy or chemical potential for the kink; as in the previous
section, we need not keep track of this.) The final picture that emerges
is Eq.(\ref{kinkpart}), with $m_{i}$ restricted to $0,\pm 1$ and 
$\beta_{eff}$ replaced by $\beta f^2.$
\vskip -0.25cm

\subsection{\noindent Relationship between the representations}

The kink antikink gas, that we derived from Eq.(\ref{energy}) by a
sequence of approximations, thus leads to a partition function that is
essentially the {\it same\/} as obtained in the phonon representation,
except that $\beta f^2$ is replaced by $4/\beta f^{2}.$ This is not 
surprising, since the continuum one-dimensional long-ranged sine-Gordon
model has such a duality.\cite{schmid} However, in the discrete version,
when the strength of the potential $V(u_i)$ in Eq.(\ref{htenergy}) tends 
to infinity, corresponding to a lattice model, the fugacity of the kinks
in the kink representation does not tend to zero. This is because 
$u_i$ can jump from one integer to another when the discrete variable $i$
is increased by unity, whereas discontinuities in $u(x)$ are not allowed
in the continuum version. 

For $\beta
=2/f^2,$ the system is at its self dual point. Of course, the duality that 
we have arrived at is only an approximate one. In particular, if one starts
from Eq.(\ref{energy}) at the self-dual $\beta=2/f^2,$ and proceeds along 
the phonon and kink routes, the resultant chemical potentials and various
sublogarithmic interaction terms are different. In Section 3, we 
shall obtain an
exact duality for a model using a restricted class of periodic potentials
$V(u)$, and a phonon spectrum that is linear in $|q|$ only for small $q.$

\subsection{\noindent Renormalization Group}

The long-wavelength limit of Eq.(\ref{phonon2}) has been studied using the
renormalization group\cite{kane}
in connection with several different problems.\cite{fisher,tsai} We
shall cite the results here without deriving them.

{}From the $|q|$ form of the propagator for small $q,$ one can see
from power counting that the $u(x)$ field is dimensionless in the
absence of loop corrections. This is similar to the Kosterlitz-Thouless
transition\cite{kosterlitzthouless}---or the sine-Gordon model---in 
two dimensional systems. Unlike
the case there, however, the singular form of the propagator here prevents
any renormalization of $u$ (or equivalently of $\beta$) to any loop order.

If the potential $V(u)$ is expanded in its harmonics, all higher
harmonics die away rapidly compared to the lowest one, and can be
neglected. Replacing $V(u)$ with $g\cos (2\pi u/f),$ one obtains a
one-dimensional long-range version of the sine-Gordon model. Calculating
one-loop corrections, one finds
$dg/dl=g(1-2/\beta f^2) + O(g^3)$ (shifting $u$ by $f/2$ shows that there
is a $g\rightarrow -g$ symmetry) and hence 
the operator $g$ is  irrelevant for
$\beta <2/f^{2},$ and relevant for $\beta >2/f^{2}.$ Although this is a
weak-coupling result, it has been argued to be true even for large $g.$
At $\beta =2/f^{2},$ by mapping the problem to the scattering from a
potential of one dimensional free fermions, it can be shown that the
behavior is not universal, and depends on $g$.

In the high temperature phase, Eq.(\ref{phonon2}) thus renormalizes to
a harmonic phonon energy. The form of the density density correlations
can be obtained from the following argument. The deviation of the density 
from the mean has a lowest
Fourier component of the form $\delta\rho(x)=\cos(2\pi f x +\theta(x)),$
where $\theta(x)$ varies more slowly than the oscillations in 
$\cos(2\pi f x).$ Comparing with the definition of the (coarse-grained)
displacement field $u(x),$ we see that this is effectively
$\cos\big[2\pi(fx+u(x))\big].$ The connected part of the 
density correlation function is then of the form 
\begin{equation} 
K_c(x-y)\sim\big\langle \delta\rho(x) \delta\rho(y)\big\rangle\sim 
\Big\langle\cos 2\pi\big(f x +u(x)\big)\cos 2\pi\big(f y + u(y)\big)\Big
\rangle.  
\label{htcorr1} 
\end{equation} 
Using the fact that $u(x)$ is a Gaussian field, this simplifies to
\begin{equation} 
K_c(x-y)\sim\cos 2\pi(fx-fy)\exp\bigg\{-2\pi^2\Big\langle
\big[u(x)-u(y)\big]^2 \Big\rangle\bigg\}\sim{\frac{{\cos 2\pi(fx-fy)}}
{{\ |x-y|^{4/\beta}}}} 
\label{htcorr2} 
\end{equation} 
where the last expression is an asymptotic result. Thus $K_c(x-y)$ is a
product of two terms: a rapidly oscillating factor, corresponding to
the periodicity of the ideal crystalline state, and a factor that decays
as a power of the separation between the points. In the low temperature
phase, where $g$ is relevant, $u(x)$ is almost always at the minima of
the potential $V(u),$ and there is long range order in the system. This
is easiest to see in the kink representation.

For $\beta>2/f^2,$ proceeding through the kink representation yields
Eq.(\ref {phonon2}), with $\beta\rightarrow 4/(\beta f^4),$ {\it i.e.\/}
in the high temperature phase. As we shall now see, the resulting {\it
irrelevance\/} of the operator $g$ implies long range order in the kink
representation. The relative phase of the local periodic structure at two
points $x$ and $y,$ compared to a reference crystalline configuration,
is $2\pi f\big\{n_k(x,y)-n_{\overline k}(x,y)\big\},$ 
where $n_k(x,y)$ and $n_{\overline k}(x,y)$ are the number of kinks and 
antikinks respectively between $x$
and $y.$ As in the phonon representation, the correlation function
is then of the form 
\begin{equation} 
K_c(x-y)\sim\bigg\langle\exp\Big[2\pi i f \big\{n_k(x,y)-
n_{\overline k}(x,y)\big\}\Big]\bigg\rangle.  
\label{ltcorr1} 
\end{equation} 
Since in the kink representation the potential $V(u)$ in
Eq.(\ref{phonon2}) with $\beta\rightarrow 4/\beta f^4$ 
is interpreted as the generating function for
the kinks and antikinks, the right hand side of Eq.(\ref{ltcorr1})
can be obtained by changing $V(u)$ in the region between $x$ and $y$
from $g\cos(2\pi u/f)$ to $g \cos(2\pi f +2\pi u/f)$ (so that $e^{\pm
2\pi i u/f}$ picks up a factor of $e^{\pm 2\pi i f}$). If the partition
function thus modified is denoted by $Z(g,g^\prime), $ and the original
partition function by $Z(g,g),$ we have 
\begin{equation} 
\bigg\langle\exp\Big[2\pi i f \big\{n_k(x,y)-
n_{\overline k}(x,y)\big\}\Big]\bigg\rangle
={\frac{{Z(g,g^\prime)} }{{Z(g,g)}}}.  
\label{ltcorr2} 
\end{equation} 
If $g$ is an irrelevant operator, this flows to a constant at long
distances under renormalization (the value of the constant depends
on the finite corrections that are removed along the course of the
renormalization flow), and there is long range order. By expanding the
right hand side of Eq.(\ref {ltcorr2}) in powers of $g$ for small $g,$
it can be seen that the correlation function decays to its long distance
limit with a power law transient rather than an exponential.

An alternative way to understand the correlations in the kink
representation is to start with Eq.(\ref{ltenergy}). At low temperatures,
the system consists of bound pairs of kinks and antikinks. The probability
that the number of kinks and antikinks between two points $x$ and $y$
separated by a large distance are not matched, is then dominated by cases
when a kink (antikink) lies just inside the interval $(x,y)$ and its
partner lies just outside. This is clearly independent of the separation
between $x$ and $y$ when the separation is large, so that $K_c(x-y)$ goes 
to a non-zero limit for large $|x-y|.$ The phase transition then 
corresponds
to an unbinding point, where the mean separation between bound pairs
diverges, {\it i.e.\/} $\int^{|x-y|} dl\,( l/l^{\beta f^2})$ diverges
for large $|x-y|.$ This occurs at $\beta=2/f^2.$ Beyond this point,
there is a proliferation of kinks and antikinks; for large separations
${\rm mod}(n_k-n_{\overline k} ,1/f)$ is equally likely to assume any
of its allowed values, and there is no long range order.

In more detail, the density density correlation function can be expressed 
in terms of the set of functions 
\begin{equation}
C^{(\nu )}(j-k)=\,\Big\langle\exp \big\{2\pi i\nu (u_{j}-u_{k})\big\}
\Big\rangle,
\label{cfndef}
\end{equation}
with $\nu=1,2\ldots.$ The preceding discussion only deals with $\nu=1$;
other values of $\nu$ give rise to corrections to the correlation function
that decay more rapidly, and therefore do not affect the leading asymptotic
behaviour. 

\section{Exact Duality in Villain-sine-Gordon Theory}

In this section we consider a particular type of sine Gordon Theory that
corresponds to a Villain approximation\cite{villain} of the cosine
function, hence the Villain sine Gordon model (VsG). The advantage of
this model is that one has an exact duality reminiscent of the Kramers
Wannier duality in the 2-d Ising model. Towards this end we begin by
considering a model for the energy in the sense of models described in
Eq.(\ref{phonon2}), a sine Gordon model given by
\begin{equation}
\beta H_{sG}=\frac{\pi \beta }{2}\sum_{q}h_{q\;}\widetilde{u}_{q}
\widetilde{u}_{-q}+\beta g\sum_{j=1,N}\big[1-\cos (2\pi u_{j}/f)\big].
\label{sgenergy}
\end{equation}
The Gaussian propagator $h_{q}$ is specified partly by giving its leading
behavior as $h_{q}=|q|+O(q^{2})$ for small q, the sum is over the $N$
wavevectors $q,$ obeying $-\pi \leq q\leq \pi $. The partition function
is obtained by writing

\[Z_{sG=}\int \Pi _{j}\,du_{j}\exp (-\beta \,H_{sG})\]
The Villain version of this model is defined by the partition function 
\begin{equation}
Z_{VsG}[\beta ,g]=\sum_{\xi _{j}=0,\pm 1,.._{{}}}\int \Pi _{j}\,du_{j}\;
\exp\bigg[-\frac{\pi \beta }{2}\sum_{q}h_{q\;}\widetilde{u}_{q}
\widetilde{u}_{-q}-\frac{2\pi ^{2}}{f^{2}}g\beta \sum_{j}(u_{j}-f\;
\xi _{j})^{2}\,\bigg]
\label{vsgpart}
\end{equation}
corresponding to a periodic function replacing the cosine in Eq.(\ref
{sgenergy}), as usual, with the correct quadratic coefficient. The VsG
model is defined by the above partition function, and the rules for
computing the correlation functions given in Eq.(\ref{cfndef}), in 
terms of which the original density density correlation function of 
Eq.(\ref{wvfn}) can be expressed. (Formal expressions for $C^{(1)}(j-k)$
for the VsG model are given in the Appendix, but the 
renormalization group arguments of the previous section are sufficient 
to obtain the qualitative behaviour.) The
evaluation of the partition function is done by two different ways, leading
to the same kink partition function, but with different parameters, and
hence the duality follows.

{\bf The first method} is similar to the phonon representation described
above, and is based on the Poisson summation formula:
\[\sum_{\xi _{j}=0,\pm 1,..}\exp\bigg[-{{2\pi^2}\over{f^2}}g\beta
(u_j-f\xi_j)^2\bigg]=\sqrt{1\over{2\pi g\beta}}\sum_{m_{j}=0,\pm 1,..}
\exp\Bigg[{{2\pi i m_j u_j}\over f}-{{m_j^2}\over{2g\beta}}\Bigg]\].
This is substituted in Eq.(\ref{vsgpart}), leading to a shifted
Gaussian in the variables $\widetilde{u}_{q}$, which can be integrated out,
yielding a product form for the partition function 
\begin{equation}
Z_{VsG}[\beta ,g]=(2\pi g \beta)^{-N/2}
Z_{gauss}\big(\beta ,\{h_{q}\}\big)\;Z_{vortex}[\beta ,g],
\label{factorizn} 
\end{equation} 
with 
\begin{equation} 
Z_{gauss}\big(\beta,\{h_{q}\}\big)=
\Pi _{q}\Bigg(\frac{2}{\beta h_{q}}\Bigg)^{1/2}. 
\label{gausspart}
\end{equation} 
With $\widetilde{m}_{q}$ defined as $1/\sqrt{N}\sum \exp (iq\,\,j)m_{j,},$ 
$Z_{vortex}$ is given by
\begin{equation} 
Z_{vortex}[\beta ,g]=\sum_{m_{j}=0,\pm 1,..}\delta _{0,\sum m_{j}}\;
\exp \Bigg[\bigg(-\frac{1}{2\beta g}\bigg)\sum m_{j^{{}}}^{2}-\frac{2\pi }
{\beta f^{2}}\sum_{q}\frac{1}{h_{q}}\widetilde{m}_{q}
\widetilde{m}_{-q}\Bigg]  
\label{vortexpart}
\end{equation} 
where the constraint $\sum m_{j}=0,$ arises from the vanishing of $h_{q}$
at small $q$.

{\bf The second method} is as follows. In Eq.(\ref{vsgpart}) we fix a
set of $\{\xi _{j}\}$ and then shift the variables $u_{j}=u_{j}^{\prime
}+f\;\xi _{j}$ . Next, we integrate out the (still Gaussian) variables
$u_{j}^{\prime }$ using 
\begin{equation}
\langle\widetilde{u}_{q}^{\prime }\widetilde{u}_{-q}^{\prime }\rangle\,=
\frac{1}{\beta\pi }\frac{1}{h_{q}+\alpha ^{-1}},
\label{alphadef0}
\end{equation}
where $\alpha$ is defined as
\begin{equation}
\alpha=\frac{f^{2}}{4\pi g},
\label{alphadef}
\end{equation}
and find a factorization 
\begin{eqnarray}
Z_{VsG}[\beta ,g] &=&Z_{gauss}\big[\beta ,\{h_{q}+\alpha
^{-1}\}\big]\;Z_{rough}[\beta ,g]  \label{partfactored} \\
Z_{rough}[\beta ,g] &=&\sum_{\xi _{j}=0,\pm 1,..}
\exp \bigg[-\frac{1}{2}\beta \pi f^{2}\sum_{q}\frac{h_{q}}
{\alpha \;h_{q}+1}\widetilde{\xi }_{q}\widetilde{\xi }_{-q}\bigg].  
\nonumber
\end{eqnarray}
The second part of the above is best seen as a roughening model with
discrete height variables $\xi _{j}=0,\pm 1,...$ interacting with a
potential that is long ranged $\sim 1/r^{2},$ since the propagator
is linear in $|q|$, for small $|q|.$ In order to proceed, we define
dual variables: $\eta _{j}=\xi _{j+1}-\xi _{j}.$ These satisfy the
condition $\sum \eta _{j}=0$ due to periodic boundary conditionss 
and have a natural
interpretation in terms of height differences in the roughening model. In
terms of Fourier components $(\widetilde{\eta }_{q},\widetilde{\;\xi
}_{q})=1/\sqrt{N}\sum \exp (iq\,\,j)(\eta _{j,}\xi _{j})\;,$ we have
$\widetilde{\eta } _{q}=(\exp (iq)-1)\widetilde{\;\xi }_{q}\;,$ and so
the roughening model becomes
\begin{equation}
Z_{rough}[\beta ,g]=\sum_{\xi _{j}=0,\pm 1,..}\exp \bigg[
-\frac{1}{4}\beta \pi f^{2}\sum_{q}\frac{h_{q}}{(\alpha \;h_{q}+1)
(1-\cos (q))}\widetilde{\eta }_{q}\widetilde{\eta }_{-q}\,\bigg].  
\label{roughening}
\end{equation}
The propagator is now again of the form $\frac{1}{|q|}$ for small $|q|,$
and hence we can hope to get a more exact equality. This motivates us
to choose the function $ h_{q}$ $\rightarrow h_{q}^{*}(\alpha )$ which
satisfies a quadratic equation:
\begin{equation}
\frac{1}{4}\pi f^{2}\frac{h_{q}^{*}}{(\alpha \;h_{q}^{*}+1)
(1-\cos (q))}=\frac{f^{4}}{4}\bigg[\frac{2\pi \alpha }{f^{2}}+\frac{2\pi}
{f^{2}\;h_{q}^{*}}\bigg].
\label{sdquad}
\end{equation}
Thus, provided $\alpha \leq \frac{1}{2}$, we have a self dual
propagator 
\begin{equation}
h_{q}^{*}={{2\big| \sin(q/2)\big|}\over{1 -2 \alpha\big|\sin(q/2)\big|}}.
\label{sdprop}
\end{equation}
It is readily seen that $h_{q}^{*}=|q|+O(q^2)$ for small $|q|$. This
choice gives 
\begin{eqnarray}
Z_{rough}[\beta ,g] &=&\sum_{\xi _{j}=0,\pm 1,..}\exp \Bigg[-\frac{1}{4}
\beta \pi f^{4}\sum_{q}\;\bigg(\frac{1}{g}+\frac{2\pi }{f^{2}\;h_{q}^{*}}\bigg)
\widetilde{\;\eta}_{q}\widetilde{\eta }_{-q}\Bigg]  
\label{roughdualized} \\
&=&Z_{vortex}\bigg[\frac{4}{\beta f^{4}},g\bigg].
\end{eqnarray}
The second equation follows from comparing the first with the definition
of the vortex partition function Eq.(\ref{vortexpart}). For the self-dual 
propagator, $h^*_q,$ we see that $Z_{rough}$ and $Z_{vortex}$ are in fact
{\it independent\/} of $g,$ and are equivalent to the model studied earlier
by Kjaer and Hilhorst.\cite{kjaer} However, the original $Z_{VsG}$ is 
still $g$ dependent. The implications of this will be discussed further
in the last section of this paper. Hence we have
the final result:
\begin{equation}
Z_{VsG}[\beta,g]=Z_{gauss}\big[\beta ,\{h_{q}^{*_{{}}}+\alpha
^{-1}\}\big]\;Z_{vortex}\bigg[\frac{4}{\beta f^{4}}\bigg].  
\label{lowtemppart}
\end{equation}
A comparison of the above with Eq.(\ref{factorizn}) provides the exact
duality relation for this choice of $h_{q}^{*_{{}}}:$
\begin{equation}
Z_{vortex}\bigg[\frac{4}{\beta f^{4}}\bigg]=Z_{vortex}[\beta]\;\Pi _{q}
\Bigg(\frac{h_{q}^{*_{{}}}+\alpha ^{-1}}{2\pi g \beta h_{q}^{*_{{}}}}
\Bigg)^{1/2}.  
\label{vortexduality}
\end{equation}

There are several comments to be made at this point. Firstly the
restriction $\alpha \leq \frac{1}{2}$ implies that the
coupling constant $g$ must be large enough in Eq.(\ref{vsgpart});
too weak a periodic potential would have large fluctuations in
$\widetilde{u}_{q}$, that are not acceptable to this relation. In
the limit of infinite $g$ the relation is particularly simple, namely
$\alpha =0$ and hence $h_{q}^{*_{{}}}=2|\sin (\frac{q}{2})|.$ In this
case the Villain approximation also would be exact for the sine-Gordon
model Eq.(\ref{sgenergy}). The series of equivalences that have been
established here can be summarized in the following diagram:.

\begin{center}
\begin{equation}
\begin{array}{ccc}
Z_{VsG}[\beta ,g] & \longleftrightarrow & Z_{vortex}[\beta] \\ 
\Updownarrow &  & \uparrow \\ 
Z_{rough}[\beta] &  & Z_{rough}\bigg[\frac{4}{\beta f^{4}}\bigg] \\ 
\downarrow &  & \Updownarrow \\ 
Z_{vortex}\bigg[\frac{4}{\beta f^{4}}\bigg] & \longleftrightarrow 
& Z_{VsG}\bigg[\frac{4}{ \beta f^{4}},g\bigg]
\end{array}
\end{equation}
\end{center}

Here the symbols $\downarrow $,$\Updownarrow$
and $\longleftrightarrow $ symbolize relations via
Eqs.(\ref{roughdualized}), (\ref{partfactored}) and (\ref {factorizn})
respectively. We see that the critical temperature of the model, if
unique, is constrained to be $\beta _{c}=2/f^{2}.$ There is of course
no guarantee that there is no other critical point, if so, they must
occur in pairs and satisfy the product condition 
$\beta _{1c}\beta _{2c}=(2/f^{2})^{2}.$ 
As mentioned earlier, it has been argued from renormalization group 
considerations that there is only a single critical point; numerical
evidence is presented in the next section.

\section{Numerical results}

The results in the previous sections have been based on perturbative
calculations in the parameter $g$ to obtain its relevance or irrelevance
under renormalization. Although it has been argued that such perturbative
considerations are in fact valid for all $g$\cite{kane} for the continuum
sine-Gordon theory, it is nevertheless useful to 
compare the results with numerical simulations, since the large $g$ regime
could be different for the continuum and discrete models.

Correlation functions for the discrete Coulomb gas on a ring were
computed numerically. Only the case of half-filling $(f={\frac{1}{2}})$
was considered. The numerics were performed by starting in an ordered
configuration, and evolving the system under Monte Carlo dynamics for
some temperature. Various values of $\beta$ were chosen, starting
from $\beta=4$ on the high temperature side, to $\beta=9$ on the
low temperature side. This range covers the freezing transition at
$\beta=2/f^2=8.$

\begin{figure}
\centerline{
\epsfysize 3.5in\epsfbox{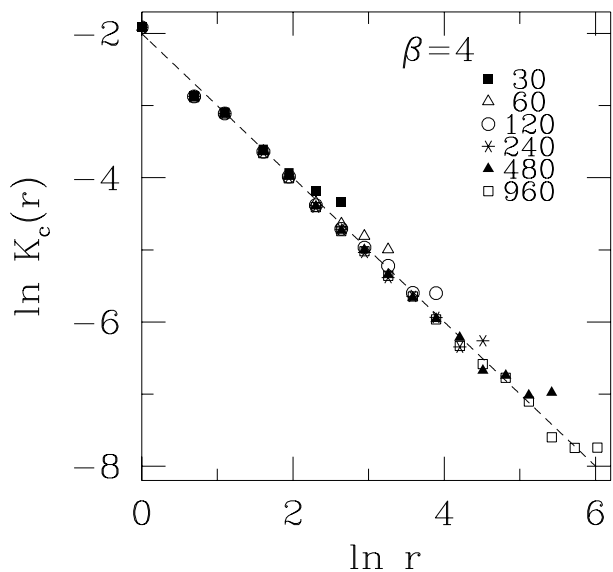}}
\vskip 0.25truecm
\caption{Log-log plot of the correlation function for $\beta=4$. System
sizes range from 30 to 960. Error bars are smaller than the symbols (except
for the last three points). The dashed
line corresponds to a power law with exponent $-1,$ as per
the exact result of Mehta.\protect\cite{mehta2} }
\label{beta4}
\end{figure}

Figure~\ref{beta4}
shows a log-log plot of the connected part of the 
correlation function, $K_c(r),$ as
a function of $r$ for $\beta=4.$ For convenience of representation, a
factor of $(-1)^r$ has been removed from $K_c(r),$ so that it represents
the deviation from perfect crystalline order. Lattice sizes of $L=30$
through 960 were used. 
\vskip -0.25cm
The correlation function is described very well
by a power law decay with an exponent 1: $K_c(r)\sim 1/r.$ This is in
agreement with the exact result.\cite{mehta2}

\begin{figure}
\centerline{
\epsfxsize 3.5in\epsfbox{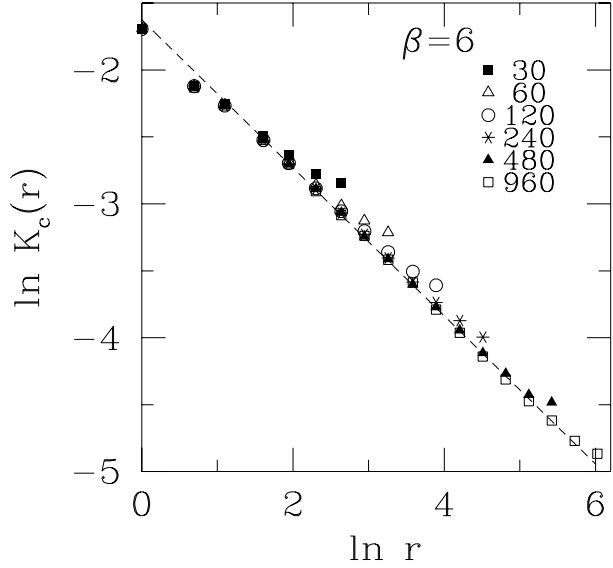}}
\vskip 0.25truecm
\caption{Log-log plot of the correlation function for $\beta=6$. System
sizes range from 30 to 960. Error bars are smaller than the symbols. 
The dashed line corresponds to a power law with the best fit exponent 
$-0.553,$ which is significantly different from the theoretical result.
This can be understood in terms of corrections to the scaling form, as 
shown in the next figure.}
\label{beta6}
\end{figure}

For any $\beta < 8,$ we expect a power law decay of $K_c(r)$ with an
exponent $4/\beta.$ However, Figure~\ref{beta6}
shows a log-log plot of $K_c(r)$
for $\beta=6;$ the apparent slope is significantly different from
$4/\beta=2/3.$ 

\begin{figure}
\centerline{
\epsfxsize 3.5in\epsfbox{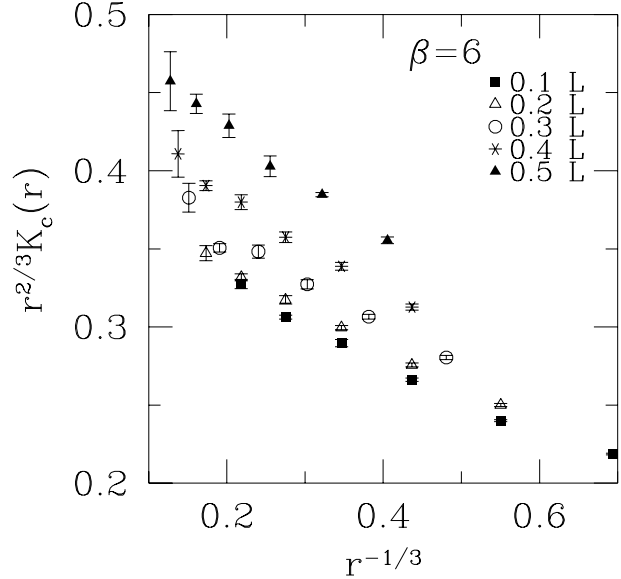}}
\vskip 0.25truecm
\caption{Correlation function for $\beta=6$, multiplied by its asymptotic
power law decay. The $x$ axis plots the $r$-dependence of the leading 
irrelevant operator, $g.$ The different symbols correspond to $r=\lambda L$ 
with different values of $\lambda.$ Apart from the smallest values 
of $r$, the data fits reasonably well to a set of straight lines. 
Note that $r$ decreases along the $x$ axis.}
\label{c2sc}
\end{figure}

One has to be careful in interpreting this result, since
as $\beta\rightarrow 8,$ the irrelevant operator $g$ renormalizes to zero
more and more slowly, so that corrections to scaling could affect the
apparent exponent over a fairly wide regime. To test whether the slope
in Figure~\ref{beta6}
can indeed be explained by leading irrelevant corrections,
we try to fit the correlation function to the scaling form
$K_c(r)\sim r^{-2/3}\hat K_c(r/L;g r^{-1/3}).$ This is the specific case
for $\beta=6$ of the general scaling form
\begin{equation}
K_c(r)\sim{\frac{1}{{r^{4/\beta}}}}\hat K_c\Big(r/L; g r^{(1-8/\beta)}\Big),
\label{irrelev}
\end{equation}
based upon the RG flow of g, namely $dg/dl= g(1-8/\beta)+O(g^3)$, with
$K_c(a,b)$ possessing a regular expansion for
\vskip -0.25cm
$a,b \sim 0$.
If $K_c(r)r^{2/3}$ is obtained for $r=\lambda L$ for fixed $\lambda$ and 
varying $L$,
the result should then be a function of $\lambda$ and $g r^{-1/3}.$ For small
$g,$ one would expect this to be a linear function of $r^{-1/3}.$ 
Figure~\ref{c2sc}
shows such a plot of $K_c(r) r^{2/3}$ as a function of $r^{-1/3}$
for various values of $\lambda.$ A set of straight lines is obtained, consistent
with the scaling prediction. 
\vskip -0.25cm
 In view of the strong dependence on $\beta$
of the dimension of the operator $g,$ it is not very useful to try
this for larger values of $\beta,$ since the range one obtains for
$L^{1-8/\beta}$ is quite limited.  Conversely, there is no sign of any
corrections to scaling for $\beta=4,$ because of the rapid decay of $g$
under renormalization.

\begin{figure}
\centerline{
\epsfxsize 3.5in\epsfbox{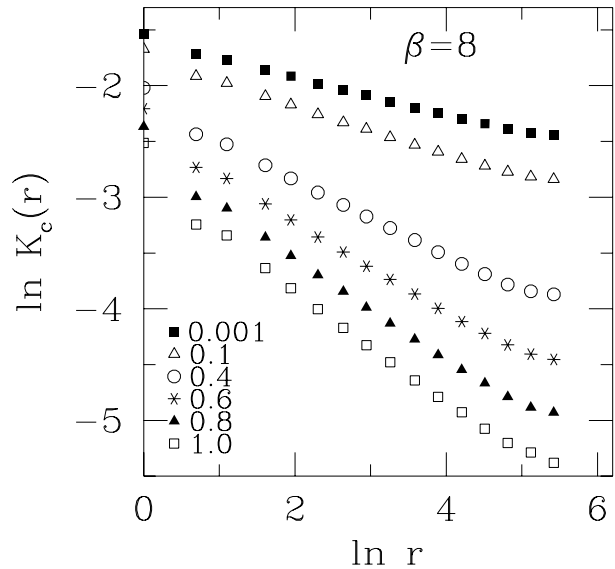}}
\vskip 0.25truecm
\caption{Log-log plot of the correlation function for $\beta=8,$ and 
system size $L=960$ showing the effect of varying the
initial $g$, a marginal operator. Error bars are smaller than the symbols.
The different curves correspond to different values of the bias, which is
the relative `Boltzmann' weight of the odd sites compared to the even 
sites. The correlation function is computed by first coarse graining the 
density, so that there are effectively 480 sites (corresponding to 240 
particles at half filling), and removing the factor of $(-1)^r.$ As the 
bias changes, the correlation function evolves smoothly, with no universality
seen even for large $r.$}
\label{beta8}
\end{figure}

At $\beta =8,$ the operator $g$ is marginal. This is not just a
perturbative result; based on the continuum model,
one expects a line of fixed points for $\beta =8,$
with continuously varying asymptotic behavior as the initial $g$
is varied. In order to check this
scenario, we performed Monte Carlo simulations on a slightly modified
version of the DCG at half filling. The lattice size was doubled,
corresponding to $f={\frac{1}{4}},$ but the particles were biased to be
on the even sites of the lattice by adding an extra potential to the odd
sites. It is clear that if the bias is infinite, the system is equivalent
to the DCG at half filling, while with zero bias, one has the DCG at
quarter filling. In general, the lattice structure can be represented as
a strong (strictly speaking, singular) potential $W(u)$ that is periodic
under $u\rightarrow u+{\frac{1}{4}},$ and an additional weak potential
$V(u)$ that is periodic under $u\rightarrow u+{\frac{1}{2}}.$ The
strength of $V(u)$ depends on the bias favoring the even sites. Since
$W(u)$ is irrelevant for $\beta <32,$ only the potential $V(u)$ affects
the asymptotics. As one adjusts the bias, which corresponds to changing
$g,$ one should see a continuous evolution in the asymptotic behavior
of $K_c(r).$ Figure~\ref{beta8}
shows that this is indeed the case for $L=960.$
\vskip -0.25cm

\begin{figure}
\centerline{
\epsfxsize 3.5in\epsfbox{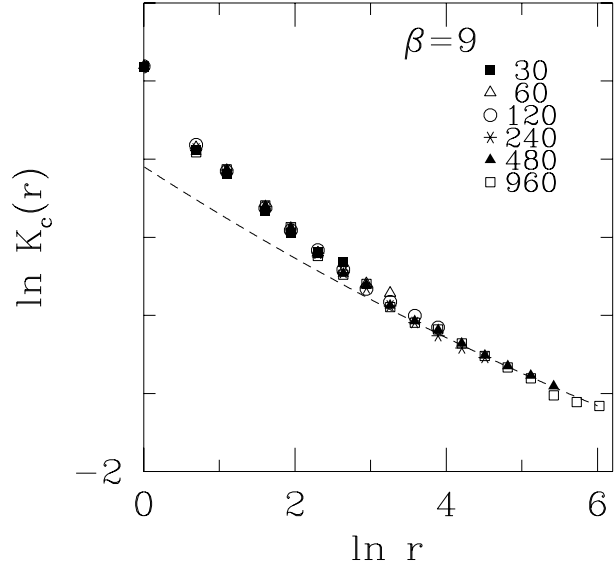}}
\vskip 0.25truecm
\caption{Log-log plot of the correlation function for $\beta=9$. System
sizes range from 30 to 960. Error bars are smaller than the symbols.
The dashed line is of the form $A+B r^{-1/8},$ which includes the leading
scaling correction to the long range order.  The parameters $A$ and $B$
are adjusted for approximately the best fit to the eye, and are both
equal to 0.1.}
\label{beta9}
\end{figure}

\begin{figure}
\centerline{
\epsfxsize 3.5in\epsfbox{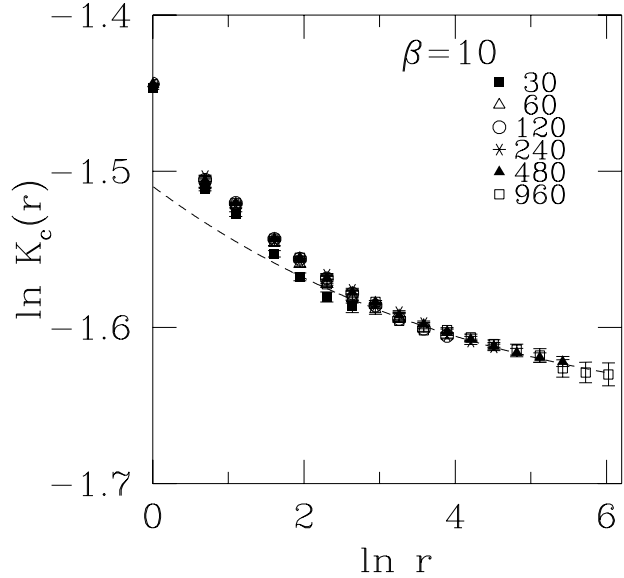}}
\vskip 0.25truecm
\caption{Log-log plot of the correlation function for $\beta=10$. System
sizes range from 30 to 960. The dashed line is of the form $A+B r^{-1/4},$
which includes the leading scaling correction to the long range order.
The parameters $A$ and $B$ are adjusted for approximately the best fit
to the eye, and are chosen to be $A=0.189$ and $B=0.032.$ With these
parameters, $C(\infty)/C(0)$ is estimated to be approximately $0.75.$}
\label{beta10}
\end{figure}

Beyond the transition, Figure~\ref{beta9} shows the correlation function
for $\beta=9,$ indicating that long-range order has set in. The slow decay
of the operator $g$ in the dual representation (since $64/\beta$ is not
much less than 8) leads to the long transients in the correlation function.
\vskip -0.25cm
By comparison, the correlation function at $\beta=10,$ shown in
Figure~\ref{beta10}, approaches the
asymptotic $r\rightarrow\infty$ limit much more rapidly.
\vskip -0.25cm
We did not increase $\beta$ beyond this, because we do not expect to see
any qualitative change in the correlation function, and because as $\beta$
is increased, it takes progressively longer for the system to equilibrate.

Thus we see that the numerical results at half filling for the density
density correlation function agree on both sides of the freezing
transition with the analytical results obtained in the previous sections.

\section{Discussion}

In this paper, we have obtained the phase diagram for the two-dimensional
Coulomb gas on a one-dimensional lattice, which emerges as a generalization
of the wavefunction of a spin $1\over 2$ Heisenberg antiferromagnet in 
one dimension. We have seen that the gas freezes into a state with long
range order below a freezing temperature $T_c= f^2/2$ that decreases as the density is
reduced. Above this temperature, the gas has quasi long range order, with 
continuously varying exponents. The dependence of the LRO on $\beta$ is thus quite dissimilar to the dependence of LRO upon $\Delta$ in the $XXZ$ model. In the latter, the isotropic point is at the brink of crystallization, whereas in this problem, the wavefunction at $\beta=4$ for the Haldane Shastry model
\cite{hal,shas} is well inside the power law phase, since crystallization only
sets in for  $\beta > 2/f^2$. On the other hand, the wavefunctions require ``umklapp'' for
$\beta > 2/f$  beyond which the kinetic energy ceases to act in a simple way
on these functions\cite{shas2}. 

The strong connections that exist to earlier
work\cite{kjaer,schmid,kane,fisher,andersonyuval} have been alluded 
to; we return to them here in more detail.
In Section 3, we considered a one parameter family of Villain sine-Gordon
models at each temperature, to obtain an exact duality transformation; the
family was parametrized by $g$. Rather surprisingly, all the models in 
the one parameter family map to the same roughening model, independent of 
$g.$ As shown in the Appendix, formal expressions for the correlation functions
of the VsG model can be obtained in terms of the roughening model,
with the parameter $g$ affecting the expressions only at short distances.
This implies that, for any inverse temperature $\beta,$ all the VsG
models (independent of $g$) flow to the same fixed point under renormalization.
Although this is reasonable at any other temperature, it is somewhat 
unexpected for $\beta=2/f^2,$
where one has a line of fixed points. Although $g$ is a marginal operator
here, the self-dual propagator  $h^*_q,$ and therefore the strength of the 
irrelevant operators, depend on $g.$ One must conclude that, within the 
one parameter family that we construct, this change in the irrelevant operators
is just sufficient to drive $g$ to the same fixed point under renormalization,
regardless of its bare value. Of course, one could obtain a duality 
mapping between VsG models with different, `conjugate', choices of $h_q,$ 
chosen to satisfy $2(1-\cos q)(\alpha + 1/h_q)=h_q^\prime/(1+\alpha^\prime
h_q^\prime).$ However, although this would allow one to access other fixed
points at $\beta=2/f^2,$ one would not have (strict) self-duality.

As mentioned earlier, the roughening model constructed in Section 3 has
been studied earlier by Kjaer and Hilhorst.\cite{kjaer} They obtained
the phase diagram we have here, and at (in our notation) $\beta=2/f^2$, 
exploit the self- duality to calculate the $\langle m_q m_{-q}\rangle$ 
correlator. 
As shown in the Appendix, the correlation function for the 
VsG model (and ultimately the Coulomb gas) is related to 
$\langle\exp i[m(x)-m(y)]\rangle.$ It is tempting to conjecture that $m(x)$
can be treated as a Gaussian variable for the long distance form of this 
correlation function, especially because using the result of Kjaer and 
Hilhorst then yields a power law decay for the VsG correlation at the 
self-dual point with the {\it same\/} exponent, $1/4,$ with two different
approaches given in Eqs.(\ref{cffirst}) and (\ref{cfsecond}). However, 
numerical simulations we have conducted for the VsG model ($f={1\over 2},
\beta=8$) do not bear this out: the numerical exponent is approximately
$1/8.$

The continuum version of the long range 1-d sine-Gordon has been studied
extensively in connection with dissipative quantum mechanics,\cite{schmid}
Luttinger liquids,\cite{kane} and the quantum Hall effect.\cite{fisher}
However, the duality transformation is slightly different from the 
discrete case.\cite{schmid} For large $g,$ $u(x)$ must be close to 
an integer everywhere; a kink consists of a rapid change of $u(x)$ from
one integer to another. If (without a lattice cutoff) the theory is 
regulated with an $m q^2$ term in the propagator for $u,$ it is clear
that $u(x)$ cannot change discontinuously. The competing effects of 
$g$ and $m$ then yield an effective kink size of $\sim\sqrt{m/g},$ and 
a kink fugacity of $\sim\exp(-\beta\sqrt{mg}).$ This is equivalent to 
a sine Gordon theory expanded in powers of the cosine interaction, provided
one chooses $g_D\sim\exp(-\beta\sqrt{mg}).$ Thus we see that the large $g$
regime maps to small $g$ under duality. This is in contrast to the discrete
model, where $u_i$ can jump from one integer to another as $i$ increases 
by 1, without any extra ({\it i.e.\/} not accounted for by $h_q$) energy
associated with the kink. Indeed, in the Villain version studied in
Section 3, a large $g$ maps to the same (large) $g$ under duality.

The renormalization group flows for the continuum model were 
obtained\cite{kane} in the small coupling constant regime. Exploiting
the duality transformation, it was possible to obtain the RG flows for 
very large coupling constants as well. It was argued that the RG flows
could be connected smoothly between these two extremities; this was 
strengthened by showing that at the self-dual `temperature', which 
corresponds to the scattering of non-interacting fermions from a barrier 
in the Luttinger liquid version, one should indeed have a line of fixed
points. 

While for a small coupling constant the discrete and continuum versions
should not differ in any physical way, this is not necessary when the 
coupling constant is large. Thus the possibility of a non-trivial 
strong coupling phase cannot be ruled out for the discrete model based
on continuum arguments. Our numerics indicate that the Coulomb gas shows
the same behaviour as the continuum model, suggesting that it is 
unlikely that there is such a strong coupling phase. Notice that,
although the RG flows are oppositely oriented for $\beta>2/f^2$ and
$\beta< 2/f^2,$ under the duality transformation of Section 3 a larger
coupling constant maps to a larger coupling constant, emphasizing the 
importance of the simultaneous change in the irrelevant operators.

A one dimensional Coulomb gas with logarithmic interactions was
considered earlier by Anderson, Yuval and Hamman\cite{ayh} in their 
study of the Kondo problem. This is equivalent to the kinks in a 
ferromagnetic Ising spin chain with long range coupling $\sim 1/r^2.$
However, the ferromagnetic nature of the underlying order forces 
charges to alternate. As shown by a real space RG 
calculation,\cite{ayh} integrating out tightly bound charge 
pairs renormalizes the strength of the logarithmic interaction 
between the remaining charges. Thus $\beta$ flows under the RG, and 
one obtains the two-dimensional Kosterlitz-Thouless\cite{kosterlitzthouless}
phase diagram. For finite coupling constant (or charge fugacity), this 
prevents one from obtaining the phase transition point exactly.

\bigskip

\centerline{\small\bf ACKNOWLEDGEMENTS}

\bigskip

We thank Chandan Dasgupta, Matthew Fisher and Peter Young for useful 
discussions. BSS
thanks the Jawaharlal Nehru Centre for Advanced Scientific Research
 at Bangalore for partial support; ON is supported 
in part by the Alfred P. Sloan Foundation.

\appendix
\section{Correlation Functions}

In this appendix, we obtain expressions for the correlation function
$C^{(1)}(j)$ within the Villain sine Gordon theory using the two
methods described in Section 3: the low and the high temperature limits. 
We begin by writing the correlation function explicitly as follows:
\begin{eqnarray}
C(k-l;\beta ,g) &=&\big\langle\exp 2\pi i(u_{k}-u_{l})\big\rangle\,
=\frac{1}{Z_{VsG}(\beta ,g)}
\sum_{\xi _{j}=0,\pm 1,..}\int \Pi _{j}\,du_{j}\exp [\Psi ],
\label{cfgen}\\
\Psi &=&-\beta \pi /2\sum h_{q}\widetilde{u}_{q}\widetilde{u}_{-q}
-\frac{2\pi ^{2}}{f^{2}}\beta g\sum_{j}(u_{j}-f\,\xi _{j})^{2}+2\pi i
(u_{k}-u_{l}).
\nonumber
\end{eqnarray}
We can proceed to evaluate this by the two methods discussed above.

{\bf The} {\bf First method: }We use the Poisson formula to trade the
integer valued variables $\,\xi _{j}$ in favor of the $m_{j}^{\prime }s$
and find
\[
C(k-l;\beta ,g)\,=\frac{1}{Z_{VsG}(\beta ,g)}\sum_{\xi _{j}=0,\pm 1,..}
\int\Pi _{j}\,du_{j}\exp\Big[-\beta \pi /2\sum h_{q}\widetilde{u}_{q}
\widetilde{u}_{-q}+2\pi i/f\sum m_{j}^{\prime }u_{j}\Big],
\]
where $m_{j}^{\prime }=m_{j}$ if $j\neq k,l$ and
$m_{k}^{\prime}=m_{k}+f,\,m_{l}^{\prime }=m_{l}-f.$ Now it is
straightforward to integrate out the Gaussian variables $u_{j}$ and in
terms of the variables $\widetilde{m}_{q}\,$defined earlier and
$\delta\widetilde{m}_{q}=\frac{f}{\sqrt{N}}(\exp (iqk)-\exp (iql)),$
we find
\begin{eqnarray}
C(k-l;\beta ,g) &=&\exp\bigg[-\frac{4\pi }{\beta N}\sum_{q}
\frac{1}{h_{q}}\big\{1-\cos(qk-ql)\big\}\bigg]\;\times
\label{cffirst} \\
&\bigg\langle&\exp -\frac{4\pi }{\beta Nf^{2}}\sum_{q}\frac{1}{h_{q}}
\widetilde{m}_{q}\,\delta \widetilde{m}_{-q}
\bigg\rangle_{\mbox{vortex}[\beta ,g]}.
\nonumber
\end{eqnarray}
The prefactor decays as a power law at all $\beta ,$ and the average in
the second term is over the vortex partition function.

{\bf The second method}: We fix the variables $\xi _{j}$ in
Eq.(\ref{cfgen}) and shift the variables
$u_{j}=u_{j}^{\prime }+f\,\xi_{j},$
and integrate over $u_{j}^{\prime }.$ We next use the difference variables
$\eta _{j}=\xi _{j+1}-\xi _{j}$, as in derivation of the roughening
model Eq.(\ref {roughening}) , and find after some manipulations
\begin{eqnarray}
C(k-l;\beta ,g) &=&\exp\bigg[-\frac{4\pi }{\beta N}
\sum_{q}\frac{1}{h_{q}+\alpha^{-1}}\big\{1-\cos (qk-ql)\big\}\bigg]\;
\times  \label{cfsecond} \\
&\bigg\langle&\exp \frac{2\pi i}{N}\sum_{q}\frac{1}{(1+\alpha h_{q})
(\exp (-iq)-1)}\widetilde{\eta }_{-q}\,\delta
\widetilde{m}_{q}\bigg\rangle_{\mbox{vortex}[4/(\beta f^{4}),g]}.
\nonumber
\end{eqnarray}
The remarkable identity of Eq.(\ref{cffirst}) and Eq.(\ref{cfsecond}) is a
consequence of the two representations of the partition function. These
are in general very hard to evaluate, the only simple situation is the
case of very low temperatures, where we can assume a dilute gas of vortex
anti vortex pairs.

\pagebreak

\end{document}